\documentclass[aps,showpacs,floatfix,twocolumn]{revtex4}
\usepackage{graphicx}      
\usepackage{float}
\newcommand{\be}{\begin{eqnarray}}
\newcommand{\ee}{\end{eqnarray}}
\newcommand{\ben}{\begin{eqnarray*}}
\newcommand{\een}{\end{eqnarray*}}
\def\beq{\begin{equation}}
\def\eeq{\end{equation}}

\begin{document}
\title{A semiclassical description of the Autocorrelations in Nuclear Masses}
\author{Antonio M. Garc\'{\i}a-Garc\'{\i}a}
\affiliation{Physics Department, Princeton University, Princeton, New Jersey 08544, USA}
\affiliation{The Abdus Salam International Centre for Theoretical
Physics, P.O.B. 586, 34100 Trieste, Italy}

\author{Jorge G. Hirsch}
\affiliation{Instituto de Ciencias Nucleares,
Universidad Nacional Aut\'o\-noma de M\'exico,
AP 70-543, 04510 M\'exico DF, Mexico}

\author{Alejandro Frank}
\affiliation{Instituto de Ciencias Nucleares,
Universidad Nacional Aut\'o\-noma de M\'exico,
AP 70-543, 04510 M\'exico DF, Mexico}

 \begin{abstract}
Nuclear mass autocorrelations are investigated as a function of the 
number of nucleons. The fluctuating part of these autocorrelations
is modeled by a parameter free model in which the nucleons are confined in a rigid sphere.
Explicit results are obtained by using periodic orbit theory. 
Despite the simplicity of the model we have found 
a remarkable quantitative agreement of the mass autocorrelations for all nuclei in the nuclear data chart.  
In order to achieve a similar degree of agreement for the nuclear masses themselves 
it is necessary to consider additional variables such as multipolar corrections to the 
spherical shape and an effective number of nucleons.
Our findings suggest that 
higher order effects like nuclear deformations or residual interactions have little relevance
in the description of the fluctuations of the nuclear autocorrelations.

\end{abstract}
\maketitle
\section{Introduction}
As a consequence of the strong nuclear interaction, the nuclear mass $M$ is not just the 
sum of the individual nucleons. The difference between these two quantities
is an indicator of the stability of a given nucleus, the larger the difference the more stable is the nucleus. 
An accurate description of this binding energy as a function 
of the number of neutrons and protons is a recurrent research topic in nuclear physics \cite{Lunn03} and  
nuclear astrophysics \cite{Rol88}.

\begin{figure}
\includegraphics[width=.95\columnwidth]{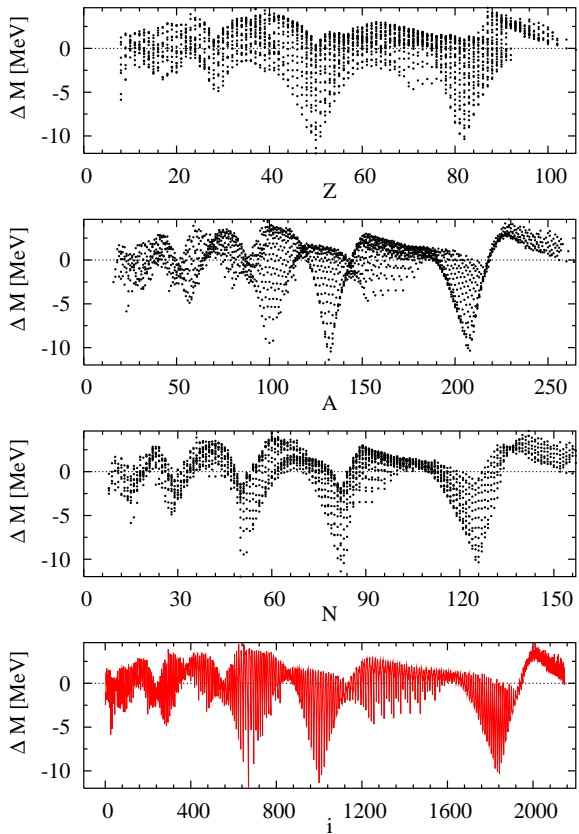}
\caption{(Color online) Mass differences plotted as function of $Z$, $A$, $N$, and as an ordered list \cite{Hir04b,Bar05}.}
\label{boustro}
\end{figure}

The semi-phenomenological liquid drop model, in which the nucleus is 
described as a very dense, charged liquid drop, is the oldest and simplest approach to this problem \cite{Bohr98}.
It provides a qualitative description of the binding energy
though it fails to capture features related to the quantum nature of the single particles 
(neutrons and protons) inside the nucleus. This is clearly observed 
  in Fig. \ref{boustro}, where we have plotted the difference 
between measured masses \cite{Aud03} and Liquid Drop Model (LDM) predictions \cite{ILDM}, 
as a function of the proton number $N$, mass number $A$, neutron number $Z$, and as an ordered list \cite{Hir04b,Bar05}.
The sharp valleys and round peaks which remain after the removal of the smooth LDM mass contribution contain information related with shell effects due to the quantum motion of the individual nucleons inside the nucleus, nuclear deformations,
and nuclear residual interactions. One of the main goals of the present paper is to further investigate the details of these corrections.
 
Most theoretical descriptions of nuclear mass models have as a starting point the general expression
\be
M = {\bar M} + \delta M
\ee
where  ${\bar M}$  is a smooth function of the number of nucleons,
usually the liquid drop mass formula. By contrast $\delta M$ is 
a fluctuating function in the number of nucleons which accounts for the  
quantum nature of protons and neutrons within the many body problem.
There is a variety of nuclear mass models in the literature, two of the most broadly utilized are  
the finite range droplet model (FRDM)  \cite{Moll95}, 
which combines the macroscopic effects with microscopic shell and pairing corrections, 
including explicit deformation effects and the Strutinsky procedure \cite{stru}, and, on the other hand, the Duflo and Zuker (DZ) \cite{Duf94} model, where the microscopic corrections 
are functions of the valence numbers of protons and neutrons. The latter is inspired in the shell model, including explicitly the diagonal two- and three-body residual interactions between valence particles and holes.

In principle the fluctuating part $\delta M$ also depends on the details of the interaction. 
However, according to Strutinsky's \cite{stru} energy theorem, the leading contribution can be evaluated within the mean field 
approximation which assumes the nucleus is composed of free nucleons confined  by a one-body potential.  
It has been shown \cite{patricio} that even a simple one-body potential, in which the nucleons are confined inside a spherical rigid sphere (spherical model from now on), with radius $R = r_0 {N_{nuc}}^{1/3}$ ($r_0 \sim 1.1 fm$ and $N_{nuc}$ the number 
of nucleons) describes qualitatively some aspects of the experimental $\delta M$.  However, for a more quantitative comparison one has to include small multipolar deformations of the 
spherical cavity and an effective number of nucleons \cite{Boh02,patricio}.
While the idea of employing a spherical well to describe the independent particle model of the nucleus is rather old \cite{Gree55}, the corresponding magic numbers, associated with the zeros of the spherical Bessel functions, are in only rough agreement with the observed nuclear shell closures, even when an effective rescaling is employed \cite{patricio}.
The spherical model has shown its best predictive power in systems with just one kind of particles, like electrons in spherical metal clusters, where shell closures are predicted in close agreement with the experimental observation \cite{Pav98}.

On the theoretical side, a clear advantage of the spherical model is that $\delta M$ can be evaluated analytically in the semiclassical limit by expressing the exact spectral density of a quantum particle in a sphere as a trace formula \cite{blo}, namely, as a sum over periodic orbits of the classical counterpart. In this way explicit expressions for $\delta M$ are available for a nucleus composed of an arbitrary number of nucleons.  
In this letter we will show that this simple spherical model is specially suitable for the  description of the autocorrelations $C(q)$ of $\delta M$ as a function of the number of particles. 
We will show that a quantitative agreement with the experimental mass autocorrelations can be obtained without any of the extensions (deformations of the sphere and an effective number of nucleons) needed for the case of the microscopic contributions to the nuclear mass. This is indeed remarkable given the simplicity of the model and the complex behavior of the nuclear many body problem.

\section{Autocorrelations}

Our object of study is the autocorrelation,
\be
\label{auto} 
C(q) = \frac{F(q)}{F(0)}\frac{N}{N-q}
\ee

with
\be
F(q) = \sum_{i}\delta M(i) \, \delta M(i+q)
\ee
where the sum runs, depending on the case, over the total number of nucleons, the neutron number $N$, or 
over a set including all possible nuclei as given by the boustrophedon list \cite{Hir04b,Bar05}. 
We shall also investigate $C(q)$ inside an isotopic chain, namely, 
we fix the number of protons $Z$ and examine the autocorrelations among all isotopes.

Autocorrelations are a useful tool in identifying relationships between elements in a list or an array. The autocorrelation of a constant distribution is also a constant distribution, and that of a pure harmonic sine or cosine distribution will also be an oscillatory distribution.
On the other hand, the autocorrelation of a random distribution is a delta function, peaked at $q=0$ plus a small random signal for any other $q$, signaling a null correlation length among the elements of the distribution. 

The fluctuating part of the nuclear mass distribution can be defined by
\be
\label{exp}
\delta M_{exp} = M_{exp} - {\bar M}_{drop}  ~,
\ee
where $M_{exp}$ is the experimental value for a certain nucleus according to the nuclear 
data chart \cite{Aud03} and ${\bar M}_{drop}$ is the prediction of the liquid drop model \cite{ILDM}.
As shown in Fig 2, the autocorrelation $C(q)$ has a well defined oscillatory behavior with 
clear maxima and minima related with the presence of shell closures, as seen in Fig. 1. 
When the oscillation amplitude decreases, the position of the first zero in $C(q)$ provides an estimate of the size of the region in the nuclear chart where the microscopic, fluctuating contributions to the nuclear masses are strongly correlated.
It will be shown that this region can include as many of 10 to 15 isotopes
or isotones, with at least 200 neighboring nuclei significantly correlated. The oscillatory behavior of $\delta M$ is closely related with the oscillations in $C(q)$. In what follows it will be shown that not only the oscillation length, but also other details of these oscillations are well described by the spherical model.

Theoretically $\delta M$ is expressed as a function of the spectral density 
$g(E) = \sum_i \delta(E-E_i) = \bar g(E) + \delta g(E)$ of the one body Hamiltonian (in our case a free fermion confined in a spherical cavity) as, 
\ben
\label{M}
\delta M = M - {\bar M}
\een
with
\be
\label{exM}
M = 2 \sum_{i=1}^{N}E_i = 2\int^{E_F}E \,g(E)\, dE\\
\ee
and 
\ben
{\bar M} = 2\int^{{\bar E}_F}E \,{\bar g}(E)  \, dE , 
\een
where $E_i$ are the eigenvalues of the one-body Hamiltonian and ${\bar g}$ and $\delta g$ are 
 the mean and fluctuating part of the spectral density respectively. 

The exact ($E_F$) and smooth (${\bar E}_F$) Fermi energies are obtained explicitly as a function of the number of particles by inversion of the following relation,
\be 
\label{N}
\frac{N_{nuc}}{2} =   \int^{E_F} g(E)\, dE = \int^{{\bar E}_F} {\bar g}(E)\, dE .
\ee
where $N_{nuc}$ is the number of nucleons (neutrons or protons) and the factor two accounts for the 
spin degeneracy.
The final expression of $\delta M$ in term of the spectral density is given by,
\ben
\label{fluc}
\delta M(N_{nuc}) = M(N_{nuc}) - {\bar M_{nuc}} \\= 2 \int^{E_F}E \,g(E)\, dE - 2\int^{{\bar E}_F}E \,{\bar g}(E)  \, dE .
\een

In order to compute analytically the autocorrelation $C(q)$ we will first evaluate $\delta M$ by using the semiclassical expression for the fluctuating part of the spectral density $\delta g(E)$ in a spherical cavity. 
We then mention how to get ${\bar E}_F$ as a function of the number of particles $N_{nuc}$. 
It is well known \cite{baduri} that, for generic cavities, the smooth part of the spectral density ${\bar g}(E)$ in three dimensions 
is given by,
\be
{\bar g}(E) = \frac{m}{2\pi^2 \hbar^2}\left [Vk + S + \frac{1}{6 \kappa}\int dS \left ( \frac{1}{R_1}+\frac{1}{R_2}\right)\right]   \label{gbar}
\ee
where $E= \hbar^2 k^2/2m$, $V$ is the volume of the cavity, $S$ is the surface, $R_1, R_2$ are the radii of curvature and $\kappa$ is the 
 scalar curvature. 

For a spherical cavity Eq. (\ref{gbar}) reduces to,
\be
{\bar g}(E) = \frac{1}{3\pi}E^{1/2}R^3 -\frac{1}{4}R^2+\frac{R}{6\pi}E^{-1/2}
\ee
In this way the mean Fermi energy is explicitly obtained as a function 
of $N_{nuc}$ by performing the integral in Eq. (\ref{N}) and then expressing ${\bar E}_F$ as a function of $N_{nuc}$.

In the following section we give a brief account of how to evaluate $\delta g(E)$ semiclassically by a trace formula involving only classical quantities. 

\section{Semiclassical evaluation of the  spectral density in a spherical cavity}  
The oscillatory part of the spectral density describes the fine structure of the spectrum. 
These oscillations are related with classical periodic orbits inside the
cavity \cite{bal} (for an introduction see \cite{baduri,patricio}),
\be
\label{osc}
\delta g(E)=\sum_{\alpha}A_{\alpha}(E)\exp(iS_{\alpha}(E)/\hbar+\nu_{\alpha}) ~,
\ee
where the index $\alpha$  labels the periodic orbits, $S_\alpha$ is the classical action
 and $\nu_{\alpha}$ is the Maslov index. 
As a general rule, the amplitude $A_\alpha(E,L)$ is a decreasing function
 of the cavity size $L$ but depends strongly on its shape. 
It increases with the degree of symmetry of the cavity. It is maximal in spherical 
 cavities and minimal in cavities with no symmetry axis. The difference (for the same volume)
 between these two limits can be of orders of magnitude.

\subsection{The spherical cavity}

The oscillating part of the spectral density of a particle 
in a spherical cavity of radius $R$ has already been analyzed in the literature \cite{blo1,blo2}. 
Below we provide a brief overview   
and refer to \cite{blo2} for an account of the details of the calculation.

For a spherical geometry the closed stationary trajectories are given by planar
regular polygons along a plane containing the diameter. 
 The length $L$ of the trajectories is given
by the simple relation $L=2pR\sin(\phi)$ where $p$ is the number of
vertexes of the polygon and $\phi=\pi t/p$ with $t$ being the number of turns
around the origin of a specific periodic orbit. Two cases must be distinguished:
Orbits with $p=2t$ corresponding with a single diameter repeated $t$ times
contribute to the density of states as, 
\be
\label{per1}
{\delta g_{D}(E)}=-\frac{1}{2\pi E_0}\sum_{t=1}\frac{1}{t}\sin(4t\sqrt{E/E_0}) ,
\ee
where $E = \frac{\hbar^2 k^2}{2m}$ and $E_0 = \frac{\hbar^2}{2mR^2}$.
For the case $p>2t$ corresponding to regular polygons the contribution to the
spectral density is given by,
\ben
\label{per}
{\delta g_{P}(E)}=\frac{1}{E_0} \left(\frac{E}{E_0}\right)^{1/4} 
\sum_{t=1} \sum_{p>2t}(-1)^t \,{\sin(2\phi)} \, 
\\ \sqrt{\frac{\sin(\phi)}{p\pi}} \,
\sin\left(\frac{3\pi}{4}+p\sin(\phi)\sqrt{E/E_0}\right).
\een
The complete expression for the fluctuating part of the spectral density is, 
\begin{equation} 
\label{oscR}
\delta g(E)=\delta g_P(E)+\delta g_D(E) ,
\end{equation}
where the first term yields the leading correction for sufficiently large cavities. 

A similar calculation can be in principle carried out for a chaotic cavity. 
In this case the spectral density can also be written in terms of classical periodic orbits by 
using the Gutzwiller trace formula. Although an explicit 
expression for the length of the periodic orbits,
equivalent to Eq. (\ref{oscR}), is not in general available in this case
it is still possible to 
estimate the amplitude of the oscillating part by using symmetry arguments. 
 This amplitude increases 
 with the symmetry of the cavity.
In cavities with 
one or several  symmetry axis
periodic orbits are 
degenerate, namely,  there exist different 
periodic orbits of the same length related by symmetry transformations.
It can be shown that the amplitude, as a function of $k$, is enhanced by a factor $(kR)^{1/2}$ \cite{bal} 
for each symmetry axis. A spherical cavity has three symmetry axis so the  
symmetry factor $S$ is proportional to $S \sim (kR)^{3/2} \gg 1$. 
The factor $R$ is a typical length of the cavity.

By contrast, chaotic cavities of the same volume have no additional 
symmetries and the symmetry factor S is unity, corresponding to the contribution
of a single unstable periodic orbit. 
Consequently finite size effects are much more important in cavities with high symmetry. 
We have now all the ingredients to compute the autocorrelation $C(q)$ as a function of the number of nucleons 
in the rigid spherical approximation for the nucleus.
 
\section{Mass autocorrelations in the nuclear spherical model. Results and comparison with experiment}

In this section we adapt our previous results to the specific case of the 
nucleus. Our aim is to evaluate the autocorrelation function $C(q)$ given in Eq. (\ref{auto}).
We now describe the smooth part of the ground state energy ${\bar M}$ by means 
of the liquid drop model. The fluctuating part $\delta M$ is computed 
 by assuming that nucleons, protons and neutrons, are confined in a spherical cavity. Obviously this is a mean field approximation
that should become better as the number of nucleons grows. For comparison with the 
experimental results we will typically remove those nuclei with $N < 30$, a region where the mean 
field approximation is not appropriate. 

In our calculations, the radius $R$ is related to the number of nucleons $N_{nuc}$ by $R=r_0{N_{nuc}}^{1/3}$ with $r_0 \sim 1.1 fm$. 
We remark that since neutrons and protons are distinguishable one has to consider these contributions separately, each one with its own Fermi energy but with the same radius.
We are now ready to write down an explicit analytical expression for $\delta M$,

\be
\label{flucn}
\delta M = 2 \int^{E_F}E \,g(E)\, dE - 2\int^{{\bar E}_F}E \,{\bar g}(E)  \, dE ,
\ee
where the spectral density $\delta g(E)$ is given by Eq. (\ref{oscR}), and
$\bar E_F$ is expressed as a function of the number of particles $N_{nuc}$ 
by solving exactly the third order equation in $\bar E_F$,
 \be
N_{nuc} = \int_{0}^{\bar E_F}{\bar g}(E) = \frac{2}{9\pi}{\epsilon}^{3/2} -\frac{\epsilon}{4}+
\frac{1}{3\pi}{\epsilon}^{1/2}
\ee
with $\epsilon = {\bar E_F}/E_0$. Finally the exact Fermi Energy $E_F$ is computed by inverting numerically Eq. (\ref{N}).
In all cases we assume a mass $m_p \sim m_n \sim 940 MeV$. 
The sum over periodic orbits in Eq. (\ref{oscR}) has a natural cutoff for scales (length of periodic orbits) such that inelastic processes which break the quantum coherence are relevant. 
In order to account for this fact we have included in the  spectral density Eq. (\ref{oscR}) 
a damping factor $k(l)=\frac{l/\xi}{\sinh(l/\xi)}$ where $l$ is the length of the periodic orbit and $\xi$ a coherence length that acts as a effective cutoff for $l \gg \xi$. 
Following the estimation of Ref.\cite{patricio} for the nuclear case
we have set $\xi \sim 5 R$. We have checked that the gross features of $C(q)$ 
do not depend on the cutoff, provided that enough periodic orbits are taken into account but 
other parameters like the amplitude of the oscillations of $C(q)$ may depend on it.
This value of the coherence length can be associated with an effective temperature \cite{baduri}
close to 1 MeV, typical of pairing energies not included in the model.

\subsection{Comparison with experimental results: $C(q)$ as a function of the number of neutrons}

We now compute $C(q)$, defined in Eq. (\ref{auto}), for a nucleus composed of $N$ neutrons and $Z$ protons with the fluctuating part of the mass given by Eq. (\ref{flucn}). 

\begin{figure*}[ht]
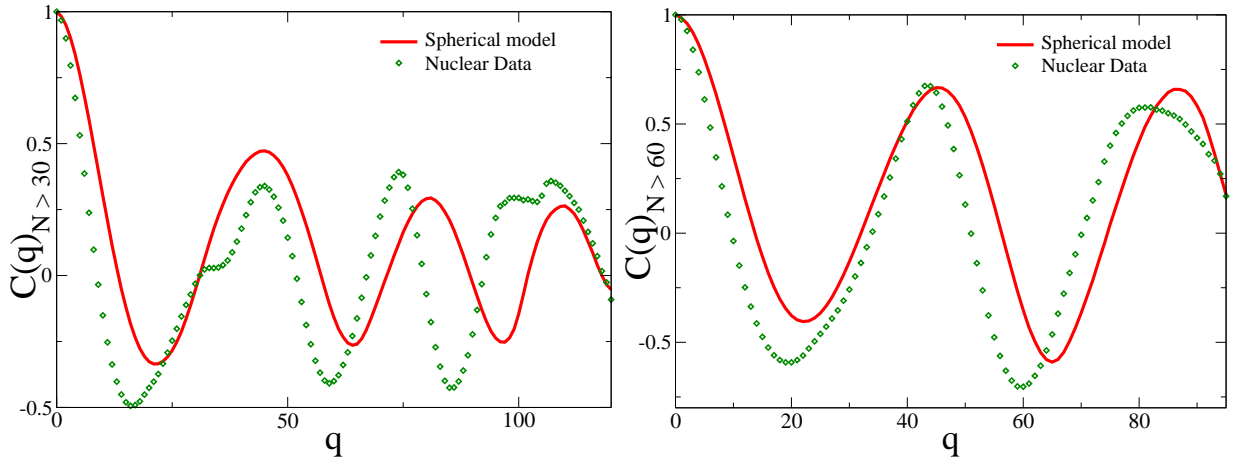

 \vspace {1cm}
 \hfill
  \begin{minipage}[t]{.45\textwidth}
     \includegraphics[width=\columnwidth,clip]{fig2a.eps}
  \end{minipage}
  \begin{minipage}[t]{.45\textwidth}
      \includegraphics[width=\columnwidth,clip]{fig2b.eps}
  \end{minipage}
     \caption{(Color online) The autocorrelation $C(q)$ as a function of $N= 30 \ldots, 154$.
(Left)  and $N= 60 \ldots, 154$ (Right). 
In both cases the agreement with the experimental results (diamond) is  quite good. 
For $N > 60$ (Right) it reproduces correctly both the amplitude of the 
 oscillations and the positions of the maxima and minima of the experimental data.}  
      \label{Nfrombous}
\end{figure*}

First we examine the autocorrelation function as a function of the total number of neutrons $N$.
We remark that predictions of our model for $C(q)$ are essentially parameter free.
Since there are many different nuclei with the same number of neutrons a proper averaging method 
is needed. In order to proceed $C(q)$ is evaluated as follows (see Fig \ref{Nfrombous} right): 
we first obtain the analytical prediction for $\delta M = \delta M(N) + \delta M(Z)$
for each of the $2140$ combinations of $N$ and $Z$, then perform an average over 
different nuclei with the same $N$ and finally compute the autocorrelation function $C(q)$.
The experimental $C(q)$ is obtained by using the same averaging procedure. 
As shown in Fig. \ref{Nfrombous}, despite the simplicity of the model, the agreement with the experimental results is quite satisfactory. It accurately reproduces both the amplitude of the oscillations and the position of the maxima and minima. 
 The agreement between theory and experiment gets better if only heavier nuclei are considered. 
This is expected due the mean field nature of the model.
The agreement between theory and experiment could be improved if, as discussed in \cite{patricio}, multipolar corrections are considered.
However we prefer to stick to our parameter-free model in order to emphasize that the main features of the autocorrelation function are related to the spherical symmetry of the problem. 
We remark that similar results are obtained if, instead of taking into account all the possible combinations of $N$ and $Z$, we make the simple assumption  
$\delta M\sim 2 \delta M(N)$,  with $r = r_0 (2N)^{1/3}$.

\begin{figure*}[ht]
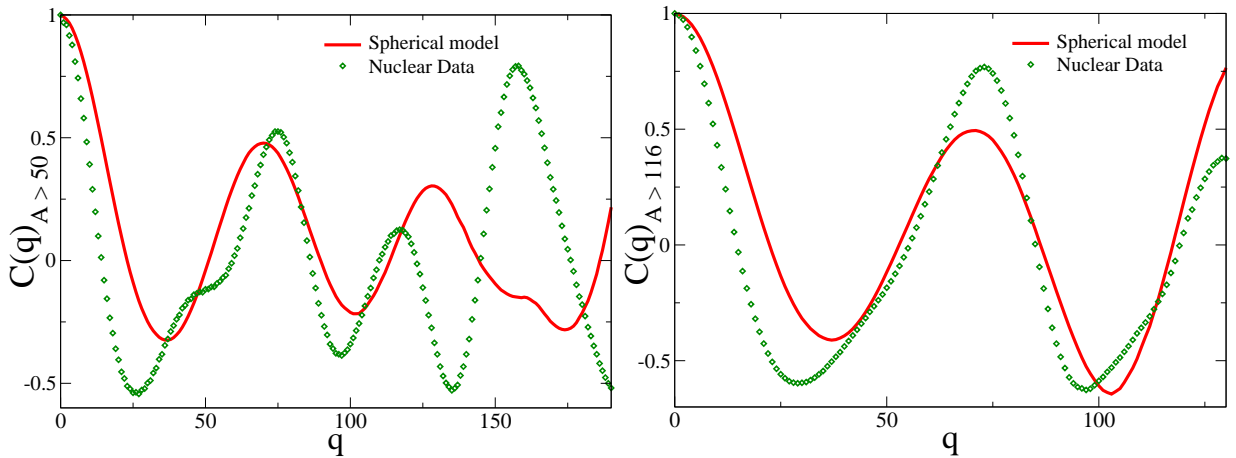

 \vspace {1cm}
 \hfill
  \begin{minipage}[t]{.45\textwidth}
      \includegraphics[width=\columnwidth,clip]{fig3a.eps}
  \end{minipage}
  \begin{minipage}[t]{.45\textwidth}
       \includegraphics[width=\columnwidth,clip]{fig3b.eps}
 \end{minipage}
     \caption{(Color online) The autocorrelation $C(q)$ as a function of $A = N+Z = 50 \ldots, 254$.
(Left)  and $A= 116 \ldots, 254$ (Right). 
In both cases the agreement with experimental results (diamond) is  quite good. 
For $A \geq 116$ (Right) reproduces correctly both the amplitude of the 
 oscillations and the positions of the maxima and minima of the experimental data.}  
      \label{Afrombous}
\end{figure*}

For the sake of completeness we have also computed $C(q)$ as a function of the 
total number of nucleons  $A = N +Z$. 
As was expected (see Fig \ref{Afrombous}) a similar degree of agreement has been found.

\subsection{Comparison with experimental results: $C(q)$ as a function of the boustrophedon ordering scheme}

\begin{figure*}[ht]
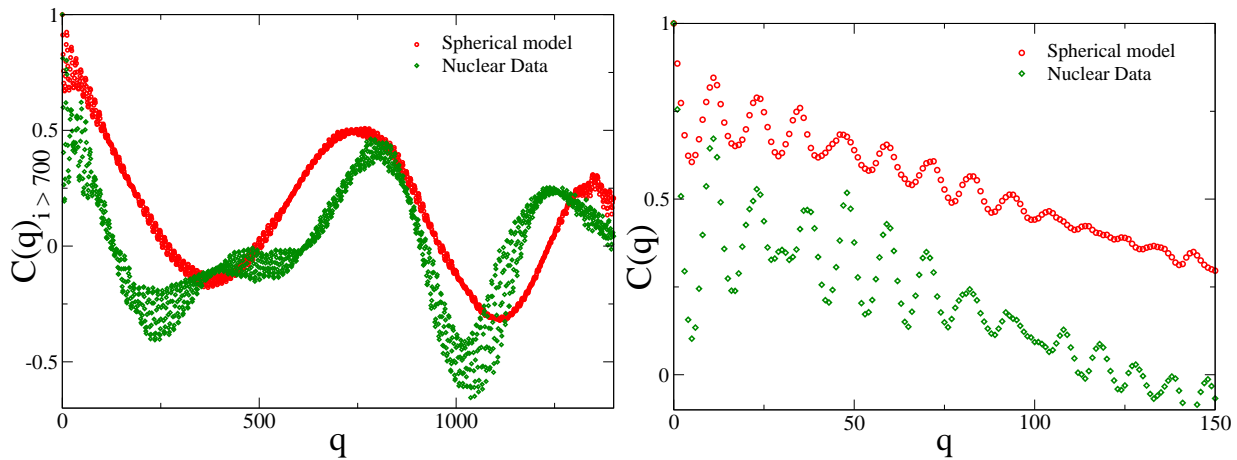

  \begin{minipage}[t]{.45\textwidth}
      \includegraphics[width=\columnwidth,clip]{fig4a.eps}
  \end{minipage}
  \begin{minipage}[t]{.45\textwidth}
      \includegraphics[width=\columnwidth,clip,angle=0]{fig4b.eps}
  \end{minipage}
\caption{(Color online) The autocorrelation $C(q)$ as a function of the order number according to the boustrophedon list. (Left) 
Both the result for the spherical model and the experimental data are obtained by considering the $2140$ possible combinations of $N$ and $Z$. 
The agreement between theory (solid line) and experimental data (diamonds) is  quite good. (Right) 
The same but now $C(q)$ is plotted only in the window $q < 150$.} 
      \label{bous}
\end{figure*}

By performing averages (for $A$ or $N$ fixed) over the nuclear data-chart we 
may be loosing valuable information about nuclear mass correlations. 
Moreover, since cuts along fixed $N$ or $A$ have a small number of nuclei, it is difficult to extract definite conclusions. 
To overcome these difficulties, we organize all nuclei with measured mass by ordering them in a
boustrophedon, namely, a 1D list composed by $2140$ entries numbered as follows: 
Even-A nuclei are ordered by increasing $N-Z$, while odd-A ones follow a decreasing value of $N-Z$. 
We have evaluated both the experimental and the analytical autocorrelation $C(q)$, Eq. (\ref{auto}), as a function of the order number of the boustrophedon.
For each $i=1,\ldots,2140$, $\delta M$ is evaluated by a specific $N$ and $Z$ combination chosen according to the above classification scheme, as we did previously, but in this case we have not performed any average.
As is shown in Fig \ref{bous} the agreement between theory and experiment 
is also quite satisfactory for this more general correlation function. 
Both the global oscillatory behavior and the more microscopic details (see right plot in Fig. \ref{bous}) are well reproduced. 

From the above extensive analysis we conclude that the main features of the nuclear mass correlations are captured by the simple spherical model. As was mentioned previously, our analytical results could be further improved 
by considering small multipole corrections to the spherical shape \cite{creagh}.
However it is remarkable that our simple spherical model can reproduce in great detail average properties of the nuclear autocorrelations.

\section{Power spectrum and integrable dynamics}
\begin{figure}[ht]
\includegraphics[width=.95\columnwidth,clip]{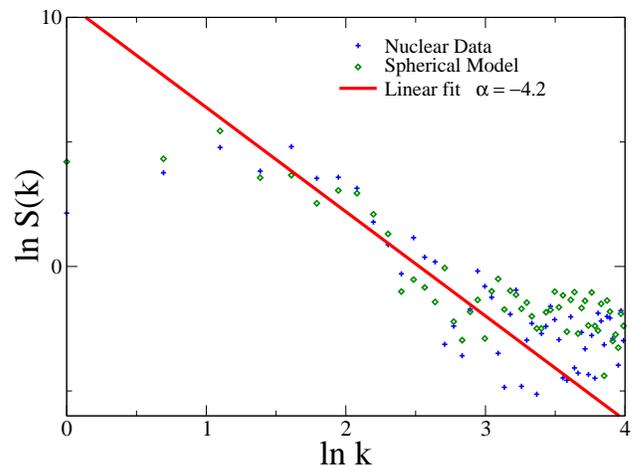}
\caption {(Color online) Power spectrum $S(k)$ of the nuclear mass fluctuation for the boustrophedon ordering. As is observed, 
the agreement between theory and experiment is very good in the low and intermediate frequency region. 
Their power-law decay with $\alpha \sim 4.2$ is close to the result
 $\alpha = 4$ predicted for classically integrable systems.}
\label{power}
\end{figure}

Finally as a further check of the validity of our results we compare the power-spectrum
associated to the nuclear mass fluctuation $\delta M(i)$ ($i=1,\ldots,  N=2140$ 
is the label of the nuclei according to the boustrophedon ordering) with the prediction
of the spherical model. 
The discrete Fourier transforms of the mass fluctuation is just, 
\begin{equation}
F(k)={\frac{1}{\sqrt{N}}}
\sum_j
{\frac{\delta M(j)}{\sigma_{\rm rms}}}
\exp\left({\frac{-2\pi ijk}{N}}\right).
\end{equation}
with the root-mean-square (rms) deviations given by
\begin{equation}
\sigma_{\rm rms}=
\left[{\frac 1 N}
\sum_{j=1}^N
\left(\delta M(j) \right)^2
\right]^{1/2},
\end{equation}
where $\delta M(j)$ 
is either the experimental or the analytical fluctuating part of the nuclear mass.
The decay of the associated power spectrum $S(k)= |F(k)|^2$
provides information about the type of dynamics of the model.
Thus it can be shown \cite{rel} that, for scales roughly in between the shortest periodic orbit 
 and the mean level spacing, a power law decay
$S(k)\sim k^{-\alpha}$ with $\alpha = 4$ corresponds to integrable classical dynamics.
In Fig \ref{power} we observe a close agreement between the power spectrum of the spherical model 
and that of the experimental nuclear masses.
Moreover the decay in the range $\sim [1,3]$, which includes frequencies  between those associated with the mean level spacing and with the shortest periodic orbit, follows a power-law 
with $\alpha \sim 4.2$, in agreement with the prediction for integrable dynamics. 
Based on these results we suggest that the power spectrum  
could be utilized as an effective test to check whether a strongly interacting many body    
system is indeed close to integrability or not. 

The power spectrum of differences between measured masses and those calculated in different
models has been studied in \cite{Bar05}. 
A gradual vanishing of the slope $\alpha$  was observed as more 
sophisticated and realistic models were utilized. 
For the most realistic models a white noise $\alpha = 0$ (all frequencies have equal weight) signal was found.  
For a detailed study of intermediate situations we refer to \cite{Hir04a}.

\section{Conclusions}

A simple semiclassical analysis, where protons and neutrons are described by free particles
bouncing elastically, back and forth inside a rigid sphere, has been shown to nearly reproduce
the autocorrelations of the differences between measured nuclear masses and those calculated using the liquid drop model. The results are remarkable, offering a different insight on the
microscopic corrections needed to describe nuclear masses with precision.
It also has been shown that it is possible to perform autocorrelation analysis of nuclear mass differences along very long chains of isotones, isotopes, isobars and other chains, a task generally considered very difficult to perform \cite{Olof04}.
While interesting in themselves, these results could also provide a theoretical explanation of
the amazing success of the two dimensional Fourier analysis, performed in the $Z-N$ space, in the description and prediction of nuclear masses \cite{Bar05b}.

AMG was supported by a Marie Curie Outgoing Fellowship, contract MOIF-CT-2005-007300. 
This work was supported in part by PAPIIT-UNAM and Conacyt-Mexico.


\begin{thebibliography}{9}
\bibitem{Lunn03} D.~Lunney, J.M.~Pearson, and C.~Thibault,
Rev.\ Mod.\ Phys. {\bf 75}, 1021 (2003).
\bibitem{Rol88} C.E. Rolfs and W.S. Rodney, {\em Cauldrons in the Cosmos},
University of Chicago Press (1988).
\bibitem{Bohr98} Aage bohr, Ben R. Mottelson, {\em Nuclear Structure} vol. 1, World Scientific, Singapore (1998); 
Peter Ring, Peter Schuck, {\em The Nuclear Many body Problem}, Springer-Verlag, New York (1980).
\bibitem{Aud03} G.~Audi, A.H.~Wapstra, and C.~Thibault, 
Nucl.\ Phys.\ A {\bf 729}, 337 (2003).
\bibitem{ILDM} S. R. Souza, et al.,
Phys. Rev. {\bf C67}, 051602(R) (2003).
\bibitem{Hir04b} J.G.~Hirsch, A.~Frank, and V.~Vel\'azquez,
Rev.\ Mex.\ F\'\i s. {\bf50} Sup 2, 40 (2004).
\bibitem{Bar05} J. Barea, A. Frank, J.G.~Hirsch, and P. van Isacker, Phys. Rev. Lett.
{\bf 94}, 102501 (2005).
\bibitem{Moll95} P. M\"oller, J.R. Nix, W.D. Myers, W.J. Swiatecki,
At. Data Nucl. Data Tables {\bf 59}, 185 (1995).
\bibitem{stru} V.M. Strutinsky, Nucl. Phys. A {\bf 122} 1 (1968).
\bibitem{Duf94} J. Duflo, Nucl. Phys. {\bf A 576}, 29 (1994);
J. Duflo and A. P. Zuker, Phys. Rev. {\bf C 52}, R23 (1995).
\bibitem{patricio} P. Leboeuf, VIII Hispalensis International Summer School, Lecture Notes in Physics, Springer-Verlag, Eds. J.M. Arias and M. Lozano; nucl-th/0406064. 
\bibitem{Boh02} O. Bohigas, P. Leboeuf, Phys. Rev. Lett. {\bf 88}, 92502 (2002).
\bibitem{Gree55} Alex E.S. Green and Kiuck Lee, Phys. Rev. {\bf 99}, 772 (1955).
\bibitem{Pav98} Nicolas Pavloff and Charles Schmit, Phys. Rev. {\bf B 58}, 4942 (1998).
\bibitem{blo} R. Balian and C. Bloch, Ann. Phys. {\bf 60}, 401 (1970).
\bibitem{baduri} M. Brack, R.K. Bhaduri, {\em Semiclassical Physics}, Addison-Wesley, New York, (1997);
~H.-J. Stockmann, {\em Quantum chaos: An introduction}, Cambridge University Press, (1999). 

\bibitem{bal} R. Balian and B. Duplantier, Ann. Phys. {\bf 104}, 300 (1977).
\bibitem{blo1} R. Balian and C. Bloch, Ann. Phys. {\bf 64}, 271 (1971).
\bibitem{blo2} R. Balian and C. Bloch, Ann. Phys. {\bf 69}, 76 (1971).
\bibitem{creagh} P. Meier, M. Brack and S.C. Creagh,  Z. Phys. D {\bf 41} 281 (1997).
\bibitem{rel} A. Rela\~no, J.M.G. G\'omez, R.A. Molina, J. Retamosa, and
E. Faleiro, Phys. Rev. Lett. {\bf 89}, 244102 (2002).
\bibitem{Hir04a} J.G.~Hirsch, V.~Vel\'azquez, and A.~Frank, Phys.\ Lett.\ B {\bf595}, 231 (2004).
\bibitem{Olof04} H. Olofsson, S. \AA berg, O. Bohigas, and P. Leboeuf, Phys. Rev. Lett. {\bf 96}, 042502 (2006).
\bibitem{Bar05b} A. Frank, J. C. L\'opez-Vieyra, J. Barea, J. G. Hirsch, V. Vel\'azquez and P.
van Isacker, Heavy Ion Physics (in press).
\end{thebibliography}
\end{document}